\documentstyle{article}
\begin{document}
\title{ Fluctuation, Dissipation and Irreversibility in Cosmology}
\author{B. L. Hu}
\date{Department of Physics, University of Maryland, College Park, Maryland
20742}
\maketitle
\vskip .5cm

\begin{center}
{\it Invited Talk given at the Conference on\\}
The Physical Origin of Time-Asymmetry,\\
{\it Huelva, Spain, Oct. 1991.\\
Proceedings eds.
J. J. Halliwell, J. Perez-Mercader and W. H. Zurek\\
Cambridge University Press, 1993}\\
(umdpp 93-55)

\vskip 1cm
{\bf Abstract}
\end{center}
We discuss the appearance of time-asymmetric behavior in
physical processes in cosmology and in the dynamics of the Universe itself.
We begin with an analysis of the nature and origin of irreversibility in
well-known physical processes such as dispersion, diffusion, dissipation
and mixing, and make the distinction between processes whose irreversibility
arises from the stipulation of special initial conditions, and those arising
from the system's interaction with a coarse-grained environment. We then
study the irreversibility associated with quantum fluctuations in cosmological
processes like particle creation and the `birth of the Universe'. We suggest
that the backreaction effect of such quantum processes can be understood as the
manifestation of a fluctuation-dissipation relation relating fluctuations
of quantum fields to dissipations in the dynamics of spacetime.
For the same reason it is shown that dissipation is bound to appear in the
dynamics of minisuperspace cosmologies. This provides a natural course for
the emergence of a cosmological and thermodynamic arrow of time and suggests a
meaningful definition of gravitational entropy.
We conclude with a discussion on the criteria for the choice of
coarse-grainings and the stability of persistent physical structures.

\newpage

\section {Introduction}

In this talk I would like to discuss the nature and origin of
irreversibility in time, or, the `arrow of time' in cosmology.
This includes physical processes in the Universe, as well as the dynamics
of the Universe itself. I will use examples from modern
cosmological theories since the sixties: i.e. the `standard' cosmology
(Peebles, 1971; Weinberg, 1972); the
chaotic (Bianchi) cosmology (Misner, 1969; Ryan and Shepley, 1975),
the inflationary cosmology
(Guth, 1981; Albrecht and Steinhardt 1982; Linde 1982),
the semiclassical cosmologies (Hu, 1982; Parker, 1982; Hartle, 1983)
and to a lesser extent, quantum cosmology (Wheeler, 1967; DeWitt, 1968;
Misner, 1972; Hartle and Hawking, 1983; Vilenkin, 1986; Halliwell 1993).
(For a layman's introduction to these theories, see, e.g., Hu, 1987.)

There are many ways irreversibility shows up in ordinary physical
processes. I shall in the first part of my talk present some well-known
examples (such as dispersion, diffusion, dissipation and phase mixing)
and discuss the nature and origin of irreversibility in them.
Distinction between dissipative processes
(which are always irreversible) and irreversible -- or `apparently'
irreversible processes (which are not necessarily dissipative) is highlighted.
I'll then use the insights gained here to discuss certain aspects of chaotic
and inflationary cosmology. In the second section
I'll discuss some not-so-well-known but important examples
involving quantum field processes such as vacuum fluctuation and
particle creation and discuss the origin of time-asymmetry in them.
This touches on basic questions like the statistical nature of the vacuum,
which underlies novel processes like the Hawking and Unruh effects
discovered in the seventies (Bekenstein, 1973, 1974; Hawking, 1975;
Davies, 1975; Unruh, 1976). In the third and
fourth sections
I shall discuss how these quantum processes influence the structure
and dynamics of the early Universe. We show that a statistical mechanical
interpretation of these so-called cosmological `backreaction' processes is
possible: they are manifestations of a fluctuation-dissipation relation
involving quantum fields. In this semiclassical theory it is the fluctuation
of the quantum field which brings about dissipation in the spacetime dynamics.
With this understanding I shall suggest some ways to examine the
notion of gravitational entropy (Penrose, 1979)--
from the entropy of gravitational fields to that of spacetimes.
As for quantum cosmology, where spacetime
and matter are both quantized, I only indicate how
the basic ideas and methods in statistical mechanics adopted above
to discuss irreversibility in cosmological processes can also
be fruitfully applied to address issues in quantum cosmology (Hu, 1991a),
but I'll shy away from extrapolations, because many
concepts remain ill-defined or ambiguous (See, e.g. Ashtekar and Stachel,
1991; Isham, 1991). On the issue of the origin of time in quantum gravity,
see, e.g., Kuchar, 1992. For a discussion of time asymmetry in quantum
cosmology, see the contributions of Halliwell, Hartle, and Hawking in
this volume.

In the conclusion I summarize the key observations. The emphasis of this
talk is to put many cosmological phenomena on the  same footing as
ordinary statistical processes and to try to understand their meaning in
terms of basic concepts in theoretical physics.
We reach the conclusion that time asymmetry in
cosmology is attributable to the same origins as those observed in
ordinary physical processes; i.,e., they are determined by the way one
stipulates the boundary conditions and initial states, the time scale of
observation in comparison with the dynamical time scale, how one decides
what the relevant  variables are and how they are
separated from the irrelevant ones,
how the irrelevant variables are coarse-grained, and what
assumptions one makes and what limits one takes in shaping the
macroscopic picture from one's imperfect knowledge of the underlying
microscopic structure and dynamics. Note that here I try only to explain
HOW time-asymmetry arises from the imposition of certain conditions or
taking certain approximations, but do not pretend to explain WHY
the Universe had to start in some particular condition, e.g., smooth, or low
gravitational entropy state according to Penrose (1979) or a state defined by
the no-boundary condition of Hartle and Hawking (1983), which can by design
hopefully `explain' time-asymmetry.
When it comes to comparing philosophical inclinations
my personal preference is that there should be no special initial state
(Misner, 1969). The challenge would be to explain the present state
of our Universe as a plausible and robust consequence of evolution from
a wide variety of arbitrary initial states.

The material in the first part of my talk is old, as old as non-equilibrium
statistical mechanics itself. The second part's results are known but
more recent--from the work of quantum field theory in curved spacetimes
applied to semiclassical cosmology. So I shall spend less time on them.
The third and fourth parts contain new results, specifically,
i) the existence of a fluctuation-dissipation
relation for dynamical quantum fields at zero-temperature (thus under
non-equilibrium conditions and detached from thermal considerations,
where most previous discussions of this relation are premised upon)
(Hu, Paz and Zhang, 1992, 1993a).
ii) the appearance of dissipative dynamics in an effective Wheeler-DeWitt
equation for the minisuperspace variables in quantum cosmology (Sinha and
Hu, 1991; Hu, Paz and Sinha, 1993). Dissipation in quantum
fields and semiclassical gravity has been discussed before (Hu, 1989,
where references to earlier work on these issues can be found).
The main emphasis in this talk is dissipation and irreversibility,
the properties of noise and fluctuation which underline many important
quantum statistical field processes are only briefly touched on.
Because of space limitation, some ideas mentioned
in my talk are not discussed here. These are:
decoherence and dissipation in quantum cosmology (for a general discussion of
the interrelation of these processes, see Hu 1991a; for specific models, see
Calzetta 1991, Calzetta and Mazzitelli, 1991, Paz and Sinha, 1991, 1992),
noise and fluctuations in semiclassical cosmology (Hu, Paz and Zhang, 1993c),
coarse-graining in spacetime and gravitational entropy (Hu, 1983, 1984; Hu,
1993; Hu and Sinha, 1993).

\section{ Irreversibility and Dissipation: Examples from well-known
processes}

Let me begin by examining a few text-book type examples of irreversible
processes to illustrate their different natures and origins.\\

\noindent {\bf A. Dispersion}

Consider the trajectory of a particle colliding with fixed hard
spheres (Ma, 1985, Sec. 26.5).  Assume that the spheres are
disks with radius $a$. The particle moves with constant
velocity $v$ and has  mean free distance $\lambda >> a$
(dilute gas approximation).
The trajectories of this particle is of course reversible in time.
However, if the incident angle of the particle on the first scattering
is changed by $\delta \theta (0)$ initially at $t=0$, then after many
collisions

\begin{equation}   | \delta \theta (t) | \ge e^{t/\tau} |\delta \theta(0)|,
     ~~ \tau = (\lambda/v)/ ln(2\lambda/a)               
\end{equation}

\noindent At sufficiently long time ,
$ |\delta \theta (t) | \approx 1$, the exit
direction is randomized by the accumulated error.  The asymmetry in the
initial and final conditions of the congruence
comes from the accumulation and magnification of the uncertainty
in the initial conditions due to the collisions, even though the dynamical
laws governing each trajactory are time-symmetric.
To trace a particular trajectory
backwards in time after a large number of collisions requires an
exponentially increasing degree of precision in the specification of the
initial condition.

This situation occurs in the inflationary cosmology, in which the scale
factor of the Universe grows rapidly $ a(t) \sim e^{Ht} $ for a certain
period of time in the early history.  Any initial small disturbance with
some functional dependence on $a(t)$ will differ exponentially in time .
Indeed this is what gives the desirable properties of inflationary cosmology
in, say, addressing the flatness and horizon problems. The apparent
irreversibility of inflation is also of this nature: not in the dynamics,
but in the inbalance of the initial and final conditions. (See, e.g., Page,
1984.)


This simple phenomenon
is amply illustrated by the many sophisticated results of modern chaotic
dynamics. There, the divergence of neighboring trajectories in phase space or
parameter space is an intrinsic property of the nonlinear Hamiltonian of
the system, not a result of coarse-graining (which is implicit in, say,
the postulate of molecular chaos in Boltzmann's treatment of gas kinetics.)
The evolution of an ensemble of such systems at some finite time from the
initial moment is often `forgetful' of their initial conditions, not
because the individual systems are insensitive to the initial conditions
(as in dissipation) but because they are overly
sensitive to them to make an accurate prediction of each system almost
impossible. It is in this sense that these systems manifest irreversibility.

Chaotic dynamics also appears in cosmology, one example is the dynamics of
the mixmaster (diagonal Bianchi Type IX) Universe
(Misner, 1969). The chaotic behavior is associated with the divergence of
trajectories which describe different world histories in the minisuperspace
(Misner, 1972) parametrized by the shape parameters ( $\beta_+, \beta_-$)
(while the deformation parameter $\alpha$ plays the role of
time in quantum cosmology). This
was pointed out by Lifshitz and Kalatnikov (1971), Barrows (1982),
Bogoiavlenskii (1985), and many others (for a recent work, see, e.g.
Berger, 1992). The collision of the `world particle' is now with the moving
`walls' arising from the anisotropic 3-curvature of the homogeneous space.
One can define quantities like `topological entropy' to measure
the trajectory instability of this nonlinear system. It is of interest to
see if the trajectories in the minisuperspace will exhibit mixing properties,
in which case all configurations of the Universe at a later time can be
equally accessed from arbitrary initial conditions.  If the trajectories
distribute unevenly in certain regions it will also be interesting to
distinguish the
set of initial conditions which give rise to such distinct behaviors.
Notice that, by contrast, in the presence of dissipative mechanisms,
as we will discuss in Example C, the trajectories in the minisuperspace
will indeed evolve to a particular region around the origin,
which corresponds to the Friedmann Universe. This signifies the
dissipation of anisotropy, a necessary condition for the implementation
of the chaotic cosmology program.\\

{\bf B. Diffusion}

Let us look at some simple examples in kinetic theory: gas expansion,
ice melting and ink drop in water. These are irreversible processes simply
because the initial states of $10^{23}$ molecules on one side of the
chamber and a piece of ice or ink drop immersed in a bath of water
are highly unlikely configurations out of all possible arrangements.
These initial conditions are states of very low entropy.
The only reason why they are special is because we arrange them to be so.
For these problems, we also know that the system-environment separation and
interaction make a difference in the outcome.
In the case of the expanding gas, e.g., for free expansion:
$ \delta S _{system} > 0, ~~ \delta S_{environ} = 0, ~~ \delta S _{tot} >0 $
whereas for isothermal expansion:
$ \delta S _{system} = - \delta S_{environ} > 0, ~~  \delta S_{tot} = 0 $.

Another important factor in determining whether a process is irreversible
is the time scale of observation compared to the dynamic time scale
of the process.  We are all familiar with the irreversible process of
an ink drop dispersing in water which
happens in a matter of seconds, but the same dye suspension put in glycerin
takes days to diffuse, and for a short duration after the initial mixing
(say, by cranking the column of glycerin with a verticle stripe of dye one way)
one can easily `unmix' them (by reversing the
direction of cranking, see, e.g., Heller, 1960). Diffusion is nevertheless an
intrinsically irreversible process.

In evolutionary cosmology, the significance of any physical processes is
evaluated in comparison with the
Hubble expansion ($H=\dot a/a$, where $a$ is the scale factor) .
Those with characteristic time scales shorter than the
Hubble time ($H^{-1}$) could have enough time to come to equilibrium with the
environment, whence one can assign some
temperature to the mixture and use thermodynamical descriptions.
Thus in the radiation-dominated era ($a \sim t^{1/2}$) one usually refers to
the temperature of the ambient photon gas as the temperature of the Universe.
However, for weakly interacting particles like neutrinos and gravitons
which are rarely collison-dominated, kinetic equations are needed
to describe their transport processes. For
quantum processes such as particle creation from the vacuum occurring
at the Planck time $t_{pl} = 10^{-43} sec$, they are
intrinsically nonequilibrium quantum processes which require a
statistical field-theoretical description. By the same token, when the
background spacetime expands very rapidly, as during the vacuum-energy-
dominated inflation epoch ($a \sim e^{Ht}$),
the ordinary pratice of describing the phase transition with
finite temperature theories may prove to be rather inadequate.
Such are the ways how
time-scales and the time dependence of the scale factor enter in cosmological
processes. Now what about the time-reversible behavior of $a(t)$ itself ?

It is often assumed that the dynamics of the Universe in the contraction
phase (say, in a closed Friedmann model) is identical with the expansion phase,
because the Einstein equation is time-reversal invariant. (Of course more
coalescing and greater inhomogeneity will appear in the contraction phase
due to the phase-space difference). One can ask: How about deflation--
Is deflation during the contracting phase just as likely to happen as
inflation in the expanding phase? The answer to this question depends not
on the dynamics, as all cosmological models based on Einstein's theory
are time-reversal invariant, but on the initial conditions.
Specifically, can the conditions conducive to these different
behaviors exist with equal likelihood in the expansion
and contraction phases for these universes? The
radiation-dominated condition responsible for  the Friedmann-class of behavior
can be assumed to hold approximately at the beginning of the contracting phase
just as in the expanding phase. However, the vacuum-dominated condition may
not be so.
This is because inflation is associated with phase transition--be it via
nucleation (`old') or spinodal decomposition (`new')-- which is not necessarily
time-symmetric. To answer this question one should analyze
the probability for vacuum energy dominance to occur as the temperature
of the Universe increases during contraction,
as the broken symmetries are restored, and as the curvature
and inhomogeneities of spacetime grow in the approach towards the big crunch.
Recent results suggest that deflation is less likely (Goldwirth 1991).\\

\noindent {\bf C. Dissipation}

There are two basic models of dissipation in non-equilibrium statistical
mechanics: the Boltzmann kinetic theory of dilute gas, and the
(Einstein-Smoluchowsky) Langevin theory of Brownian motion. Each invokes
a different set of concepts, and even their relation is illustrative.
In kinetic theory, the equations governing the $n$-particle distribution
functions (the BBGKY hierarchy) preserve the full information of an $n$
particle system.
It is the truncation of this hierarchy, a procedure justified
when one is only interested in the behavior of the low-order correlation
(usually the one-particle distribution) functions, that dissipation appears.
It is in ignoring (more often restricted by the precision of one's observation
than by choice) the information contained
in the higher-order correlations which brings about dissipation and
irreversibility in the dynamics of the lower-order correlations.
(Zwanzig, 1961; Prigogine, 1962; Balescu, 1975;
de Groot, van Leeuven and van Weert, 1980; Calzetta and Hu, 1988).
For the Brownian motion
problem modeled, say, by a set of coupled oscillators with one oscillator
(mass $M$) picked out as the Brownian particle and the rest (with mass $m$)
serving as the bath (Rubin, 1960; Ford, Kac and Mazur, 1963;
Feynman and Vernon, 1963; Caldeira and Leggett, 1983).
Dissipation in the dynamics of the system
arises from ignoring details of the bath variables
but only keeping their averaged effect on the system
(this also brings about a renormalization of the mass
and the natural frequency of the Brownian particle).
Usually one assumes $M>>m$  and weak coupling of the system and the bath
to simplify calculations. The effect of the bath can be summarized by
its spectral density function, which is not unique to any
particular bath. In both of these models, as well as in more general cases,
the following conditions are essential for the appearance of dissipation
(Hu, 1989,1990; Calzetta, 1990, 1991):

\noindent a) {\it system-environment separation}. This split depends
on what one is interested in: it could be the slow variables,
the low modes, the low order correlations,
the mean fields; or what one is restricted to: the local domain, the late
history, the low energy, the asympototic region, outside the event horizon,
inside the particle horizon, etc. We shall bring up this issue again
at the end of this talk.

\noindent b) {\it coupling}. The environment must have many degrees
of freedom to share and spread the information from the system;
its coupling with the system must be effective
in the transfer of information (e.g., non-adiabatic) and
the response of the coarse-grained environment must be `sluggish'
(sufficiently degrading) in that it will only react to the system in a
non-systematic and retarded way.

\noindent c) {\it coarse-graining}. One must ignore or down-grade the full
information in the
environmental variables to see dissipation appearing in the dynamics of
the open system. (The time of observation enters also, in that it has to
be greater than the interaction time of the consitituents but shorter
than the recurrence time in the environment).
Coarse-graining can be the truncation of a correlation hierarchy,
the averaging of the higher modes, the `integrating out' of the fluctuation
fields, or the tracing of a density matrix (discarding phase informations).
See the last section for more discussions on this point.

\noindent d) {\it initial conditions}. Whereas a dissipative system
is generally insensitive to the initial conditions
in that for a wide range of initial states
dissipation can drive the system to the same final (equilibrium)
state, the process is nevertheless possible only if the initial state is
off-equilibrium. The process manifests irreversibility also because the initial
time is singled out as a special reference point when the system is prepared
in that particular initial state.
Thus in this weaker sense, dissipation is also a consequence of
specially prescribed initial conditions.
\footnote {Note the distinction between these cases: If one defines $t_0$
as the time when a dissipative dynamics begins and $t_1$ as when it ends,
then the dynamics from $t_0$ to $-t$ is exactly the same as from $t_0$ to
$t$, i.e., the system variable at $-t_1$ is the same as at $t_1$.
This is expected because of the special role assigned to
$t_0$ in the dynamics with respect to which there is time-reversal
invariance, but it is not what is usually meant by irreversibility
in a dissipative dynamics. The arrow of time there is defined as the
direction of increase of entropy and irreversibility refers to the
inequivalence of the results obtained by reversing $t_0$ and $t_1$
(or, for that matter reversing $t_0$ and $-t_1$), but not between
$t_1$ and $-t_1$. The time-reversal
invariance of the H-theorem has the same meaning.}

While the original combined system and environment still preserve
the unitarity of motion, and its entropy remains constant in time,
under these approximations, the subsystem becomes an open system, the entropy
of the open system (constructed from the reduced density matrix by tracing out
the environmental variables) increases in time, and irreversibility appears in
its dynamics.

Both irreversible (but non-dissipative) processes and dissipative
(and irreversible) processes depend on the stipulation of special
initial conditions. The difference is that the former depends sensitively
so, the latter insensitively. Dissipative processes invole coarse-graining
while non-dissipative processes do not. However, both
type of irreversible processes (Case B and C) can entail entropy generation
(even in Case A one can associate some mathematical entropy to describe the
divergence of the trajectories).
Irreversible processes described by the second law is what usually
defines the thermodynamic arrow of time.

In the context of dissipative processes, it is important to distinguish
dissipation from phase mixing, which, though sometimes called damping
(e.g. Landau damping) and has the appearance of an irreversible process,
is actually reversible.\\

\noindent {\bf D. Phase Mixing}

Two well-known effects fall under this category: Landau damping and spin echo
(e.g., Balescu, 1975, Sec.12.2; Ma, 1985, Sec. 24.3). Let us examine the
first example.
In the lowest order truncation of the BBGKY hierarchy
valid for the description of dilute gases, the  Liouvillian
operator $L$ acting on the one-particle distribution function $f_1(r_1,p_1,t)$
is driven by a collision integral involving a two-particle distribution
function $f_2(r_1,p_1,r_2,p_2,t)$:
\begin{equation}
 [\frac{\partial}{\partial t} + \frac{{\bf p}_1}{m} \cdot {\bf \nabla}_{r_1}
+ {\bf F}(r_1) \cdot {\bf \nabla}_{p_1} \big]f_1({\bf r}_1,{\bf p}_1,t) =
\big(\frac{\partial f_1}{\partial t}\big)_{coll}          
\end{equation}
$$
(\frac{\partial f_1}{\partial t}) _{coll} =
(\frac{N}{V}) \int [{\bf \nabla}_{r_1}V({\bf r}_1,
{\bf r}_2)] \cdot {\bf \nabla}_{p_1}f_2 ({\bf r}_1,{\bf p}_1,{\bf r}_2,
{\bf p}_2,t)]d^3r_2 d^3p_2
$$
The molecular chaos ansatz assumes an initial uncorrelated state between two
particles (a factorizable condition): $f_2(1,2)=f_1(1)f_1(2)$,
i.e., that the probability of finding particle 1 at $(r_1,p_1,t)$ and
particle 2 at $(r_2,p_2,t)$ at the same time t is equal to the product of
the single particle probabilities. When this condition is assumed to hold
initially and finally in a collision processes, (but the two collision partners
are assumed to be correlated within the short range of the interaction force),
one gets the Boltzmann equation. However, for long-ranged forces
such as the Coulomb force in a dilute plasma gas where close encounters
and collisions are rare, the factorizable condition can be assumed to
hold throughout. In such cases the kinetic equation becomes a Vlasov (or
collisionless-Boltzmann) Equation: (e.g., Balescu, 1975; Kreuzer, 1981)
\begin{equation}
\{\frac{\partial}{\partial t} + \frac{{\bf p}_1}{m} {\bf \cdot \nabla}_{r_1}
+ [{\bf F}({\bf r}_1) - {\bf \nabla}_{r_1} \bar \Phi ({\bf r}_1,t)] {\bf
\cdot \nabla}_{p_1} \} f_1({\bf r}_1, {\bf p}_1,t) = 0       
\end{equation}
Here
\begin{equation}\bar \Phi({\bf r}_1,t)= (\frac{N}{V}) \int V({\bf r}_1,{\bf
r}_2
   )f_1
({\bf r}_2,{\bf p}_2,t)d^3r_2d^3p_2                          
\end{equation}
is the mean field potential experienced by any one particle produced
by all other particles. It is determined by the density excess
over the equilibrium value. The effect of the mean field potential is
similar to the Debye-Huckel screening in dilute electrolyte systems.
The dependence on $f_1$ makes the Vlasov equation nonlinear:
Equations (3) and (4) have to be solved in a self-consistent way.
This is analogous to the Hartree approximation in many-body theory.
Note that the Vlasov equation which has a form depicting free streaming
is time-reversal invariant: The Vlasov term accounting for
the effect of the averaged field does not bring about entropy generation.
This mean-field approximation in kinetic theory,  which yields a
unitary evolution of reversible dynamics, is, however, only valid for times
short compared to the relaxation time of the system in its approach
to equilbrium. This relaxation time is associated with the collision-induced
dissipation process.

Landau damping in the collective local charge oscillations described by
the Vlasov equation is only an apparently irreversible processes.
The appearance of `damping' depends critically on some stipulated special
initial conditions. This damping is different from the dissipation process
discussed in Case C,
in that the latter has an intrinsic time scale but not the former, and that
while dissipation depends only weakly on the initial conditions, mixing
is very sensitive to the initial conditions. A more appropriate
name for these processes is `phase mixing' (Balescu 1975). Spin echo
is a somewhat different example of phase mixing.


{}From all of the above examples we see that irreversibility and dissipation
invole very different processes. The effect of interaction, the role of
coarse-graining, the choice of time-scales,
and the specification of initial conditions
in these proceses can give rise to very different results.
In the next section we shall use these examples to illustrate the statistical
properties of quantum field processes in the early Universe.


\section{ Fluctuations and Irreversibility: Examples from cosmological
 particle creation}

     We see in the above the many origins of irreversibility and the
distinction between dissipative and irreversible processes. Let us continue
exploring these conceptual issues now by adding an additional dimension,
fluctuations -- both quantum and thermal fluctuations. These refer to
statistical variations from the mean --
the vacuum or the background field in the case of quantum fluctuations,
the equilibrium state or the mean field in the case of thermal fluctuations.
Only quantum fluctuations exist at zero temperature.
(Their relation is an interesting issue in itself, involving
the viability of background separation, applicability of the notion of
ensembles, and the definition of classicality, to name just a few. See,
Hu and Zhang, 1992, 1993; Calzetta and Hu, 1993b)
Processes involving fluctuations play important roles in cosmology. Examples
are: Fluctuations in background spacetimes induce density contrasts as
seeds for galaxy formation (Hu, Paz and Zhang, 1993b);
parametric amplification of vacuum fluctuations leads to particle creation
in the early Universe (Parker, 1969; Zel'dovich, 1970);
fluctuations of quantum fields bring about
phase transitions in the inflationary cosmology (Guth, 1981; Sato, 1981;
Linde, 1982; Albrecht and Steinhardt, 1982); thermal fluctuation
(noise)-induced phase transitions (the Kramer process).
Even the creation of the Universe (and its babies!) has been attributed
to fluctuations of spacetime geometry and topology.
(Vilenkin, 1986; Coleman et al, 1991)

For a description of fluctuations, at least two factors,
the number of samples taken and the time of observation, usually enter into
the consideration: For $N$ samples of a system in equilibrium,
the fluctuations of physical quantities associated with
the system are of the magnitude ~$N^{-1/2}$,
and can be made arbitrarily small by making $N$ large. Thus in taking the
thermodynamic limit of the system, i.e., letting $N$ and $V$ large but keeping
$N/V$ constant, or, by looking at the system at longer time spans,
the occurance of large fluctuations are statistically suppressed.
The former operation forfeits the Poincare
recurrence, while the latter operation (made equivalent to averaging
over a large number of copies) assumes the validity of ergodicity. By contrast,
for finite nonequilibrium systems, large fluctuations can arise more readily.
Because
non-equilibrium systems have intrinsic time-scales, one cannot hope to get
an ensemble-averaged suppression by taking a long enough waiting time,
as in the equilibrium cases.
As for the issues of time-reversibility of events involving fluctuations,
although the appearance of a fluctuation and its disappearance are
time-symmetric, the set-up of problems involving
fluctuations is often such that the chronicle of interesting events starts
at the time when the fluctuation first comes into existence,
or becomes eventful. This imparts the subsequent history an apparent
arrow of time. Thus we talk
about the `beginning' of a new phase, or the `genesis' of the Universe, as if
time only exists after that particular moment.

Irreversibility and thermal fluctuations are studied in many textbooks
of non-equilibrium statistical mechanics. Here I want to focus
on the statistical properties of vacuum fluctuations, especially
in cosmological processes involving vacuum fluctuations.
Let us first analyze entropy-generation and irreversibility in a simple
but basic process, particle creation from the vacuum.

Pair creation involves the spontaneous or stimulated release of energy in the
amount of the threshold or above from the vacuum or from existing particles.
Note that the mechanism according to the basic physical laws is
time-symmetric. Thus, given equal initial and final conditions, pair
annihilation should be equally probable. However,
the initial condition is usually arranged differently
from the final conditions, and this is where the problem arises.
It is easier for a pair to be created than for them
to annihilate, because only particles-antiparticle pairs with $\pm k$
can do it and the two have to be brought together at the same point
in spacetime for this to happen.
(This is what is usually refered to as the phase space factor difference).

One of the reasons for our interest in vacuum
particle creation processes is to try to get a handle on the nature of the
ubiquitous, omnipotent, but mysterious and often ambiguous entity called
the vacuum. Note that by comparison with the particles
it creates, which carry precise and reproducible information content,
the vacuum understood in a naive way contains little information.
However, the vacuum is far more complex than a simple `nothing'.
It is made to play many different roles and perform many difficult tasks:
The vacuum is every rich man's garbage dump (witness all the divergences)
and every poor man's Messiah (``The Universe is a free lunch", Guth, 1981)
It is far from devoid of information, because everything can in principle
be obtained from it, given some viable mechanism (e.g., pair production)
and some luck (probability and stochasticity). Therefore the mechanisms which
transform the vacuum into physical reality is of special interest.
It is for this reason that some understanding of the statistical properties
of the vacuum is essential to launching
the adventurous but noble quest to `get everything from nothing', otherwise
known as Don Quixote's `free lunch'.

Cosmological particle creation adds into consideration an additional
factor of the influence of  background spacetimes on the vacuum
(Parker, 1969). We shall
look at just the dynamical effects here but not those effects associated with
the global structures of  spacetime such as the event horizon (Hawking, 1975).
We have  in earlier work analyzed the problem of
entropy generation from cosmological
particle creation and interaction processes. Let us try
to understand the different nature of irreversibility in these processes.

Assumming that at an initial time $t_0$ the system is in a mixed state
described by a density matrix $\rho$ which is diagonal in the representation
whose basis consists of the eigenstates of the number operators at $t_0$,
then the number of particles in mode $k$ in a unit volume at a later time $t$
is given by (Parker, 1969)
\begin{equation}
<N_k(t)> = | \beta_k(t)|^2 +  a_k <N_k(t_0)>    
\end{equation}
where $\beta_k$ is the Bogolubov coefficient measuring the mixing of the
positive and negative frequency components, and
$a_k = 1+ 2 |\beta_k(t)|^2$ is the parametric amplification
factor for mode $k$. The two parts in this expression can be understood as
the parametric amplification of vacuum fluctuations and that of particles
already present in mode $k$. The first part (spontaneous creation) always
increases while the second part (induced creation) can increase or decrease
depending on the correlation and phase relation of the initial state and
on whether the particle is a boson or a fermion.

Are these processes time-asymmetric? Is there entropy generation in a
vacuum particle creation process? The search for an answer to these
seemingly simple questions teaches us something interesting.
Let us separate the time-asymmetry question into two parts: one referring to
the
time-reversed process of pair annihilation, the other referring to the
probability of particle creation in the Universe's contraction phase.

Assume the Universe is in the expansion phase. Consider first the
more complicated but conceptually easier case of particle creation with
interaction. If we measure only the one-particle distribution,
the entropy function constructed
from the reduced density matrix will under general conditions
(asumming bosons with initial state an eigenstate of the number operator)
increase. (For details see Hu and Kandrup, 1987).
The primary reason is that
one has ignored the information in the higher-order correlation functions.
The presence of interaction is such
that even if one starts with an initial state with no correlation between
the relevant and irrelavant variables,
interaction can change the correlations and
bring about entropy generation.  This case is similar in nature to our example
above of dissipations in an interacting gas. These dissipative processes
are irreversible, and their outcomes usually do not depend or depend
only weakly on the initial conditions.

The other case of particle creation from the vacuum with no interaction is
more subtle (Hu and Pavon, 1986; Kandrup, 1988). On the one hand we know that
both the initial vacuum and the final particle
pair are in a pure state, so there cannot be any entropy generation. On the
other hand we clearly see an increase of particles in time, and one is tempted
to use the particle number as a measure of entropy and conclude that entropy is
generated in the process of particle creation.
(Indeed, in the thermodynamic approximation, $ S \sim N^3$, but this relation
is only valid for collision-dominated gas, which assumes interaction, from
which
entropy generation is expected). The resolution of this paradox lies in the
fact that usually in calculating particle creation one works in a Fock
space representation where the initial state (e.g., the vacuum or the
thermal state) is assumed to be an eigenstate of the number operator
($N$-representation). However, an uncertainty relation exists
between the number and the phase information.
It is at the sacrifice of the phase information that
one sees an increase of the number in time.
Had one chosen the initial
state to be of definite phase ($P$-representation), particle number will not
be monotonically increasing. Therefore it is only for the customary
choice of an eigenstate of the number operator as the initial state that the
non-dissipative process of  particle creation
with no interaction appears to be irreversible.
As in the case of phase mixing in Example D above, this apparent
`irreversibity' is also highly sensitive to the choice of the initial state.

Now consider the situation where these processes take place at the
contraction stage of the Universe and ask the question whether they will
take place with the same probability.
Let us take the simplest case of cosmological particle
creation, assuming that the {\it in}-vacuum and the {\it out}-vacuum
are well defined
(e.g., statically-bounded dynamics, or work with some conformal-vacuum)
and symmetric. Since
the Bogolubov transformations which relate a set of creation and annihilation
operators at one time to another is time-reversal invariant, the process
should be time-symmetric. That is, one should expect to see particle creation
just as likely to happen in the contraction phase. However, except for
steady state models, cosmological
conditions are not symmetric between the {\it in} and the {\it out}
states in the expanding and the contracting phases. In the expanding phase,
the {\it in}-state for particle creation processes of any cosmological
significance is usually taken to be at the singularity (`big bang')
or at least around the Planck time, while the {\it out}-state is defined
at late times before recontraction when curvature and field effects
are weak. There is asymmetry in the {\it in} and {\it out} states
between the expanding
and the contracting phases which affects the production rates. Despite
these differences, there is entropy generation associated with particle
creation and interaction in both the expanding and the contracting phases.
Thus the thermodynamic arrow of time defined by the direction of entropy
increase will see no change at the turnaround point. To the extent that
the thermodynamic arrow of time can be traced to be the root of many other
arrows of time (including the psychological), entropy generation in
particle creation can play a fundamental role in the problem of time-asymmetry.

We see in the above cosmological examples the workings of the differences
between irreversible and dissipative processes as manifested in vacuum
fluctuations and particle creation. We shall see next how these processes
can affect the dynamics of the early Universe, and manifest as a
relation between fluctuation in the quantum fields and dissipation
in the dynamics of spacetime.

\section { Fluctuation and Dissipation: Example from cosmological
  backreaction processes}

     Cosmological particle creation comes from the amplification of vacuum
fluctuations by the dynamics of the background spacetime. It is the
transformation of a microscopic random process into macroscopic proportions.
At late times like today's Universe this process is rather insignificant
(Parker, 1969). However, near the Planck time ($t_{pl} \sim 10^{-43} sec$ from
the Big Bang),
for non-conformal fields, or for non-conformally flat universes, production
of particles might have been
so copious that they could have exerted a strong influence on the
dynamics of the early Universe (Zel'dovich 1970).
In particular, anisotropies in
the early Universe can be dissipated away in fractions of $t_{Pl}$
(Zel'dovich and Starobinsky, 1971; Hu and Parker, 1978; Hartle and Hu, 1980).
Backreaction processes like these have been studied extensively for
cosmological (origin of isotropy in the Universe), philosophical (chaotic
cosmology program), and theoretical (quantum to classical transition)
inquiries. Here we'd like to view it as an example of the fluctuation-
dissipation relation relating the fluctuations of the vacuum to the dissipative
dynamics of the Universe (Hu, 1989). Take for example a massless conformal
scalar field in an anisotropic, homogeneous Bianchi Type-1 Universe with
line element
\begin{equation}
ds^2 = a^2(\eta)[ d \eta^2 - \sum^{3}_{i,j=1} e^{2 \beta_{ij}(\eta)} dx_i
dx_j].
\end{equation}
The equation of motion for the anisotropic expansion rates  $q_{ij} \equiv
\beta_{ij}'\equiv d \beta_{ij} / d \eta $
calculated in the Schwinger (1961) - Keldysh (1964) (or closed
time-path, or {\it in-in}) formalism is given by (Calzetta and Hu, 1987)
\begin{equation}
\frac{d}{d\eta} (Mq'_{ij}) + 3(2880\pi^2)^{-1} Kq'_{ij} + kq_{ij} = c_{ij},
\end{equation}
where  $c_{ij}$ is a constant measuring the initial anisotropy,
$M$ and $k$ are funtions of $a, a'$ and $a''$ .
The  nonlocal kernel $K(\eta -\eta')$
\begin{equation}
Kq'_{ij} = \int^{\eta}_{- \infty}d\eta' (\frac{d^3}{d\eta^3}
q'_{ij})ln (\eta - \eta').                                         
\end{equation}
linking the `velocities' $q'_{ij}$ at different times gives a non-local
viscosity function $\gamma $ (in Fourier space)
\begin{equation}
\gamma (\omega) = \frac{\pi}{60(4\pi)^2} \mid \omega \mid^3.        
\end{equation}
which is responsible for the dissipation of anisotropy in the background
dynamics. The energy density dissipated in the background dynamics is shown
to be exacly equal to the energy density of the particles created:
\begin{equation}
\rho(particle ~creation) = \rho (anisotropy ~dissipation)            
\end{equation}
This relation, as we pointed out earlier (Hu, 1989),
embodies the fluctuation-dissipation
relation in the cosmological context, but does not yet have the correct form
(F-D relation for black holes and
de Sitter spacetimes have been proposed by Candelas and Sciama, 1977;
Sorkin, 1986; and by Mottola, 1986 respectively).

Notice that velocity $\beta'$ enters in the equation of motion (7) instead of
displacements $\beta$. This is because the coupling between the field and the
background dynamics via the Laplace-Beltrami operator is of a derivative kind.
This equation is in the
form of a Langevin equation, except for the absence of explicit random forces.
This is because in the above example we worked with pure states to begin with
and there is no explicit coarse-graining of the environment fields.
Technically the ordinary effective action
which takes into account the averaged effect of quantum fluctuations can be
generalized to a coarsed-grained one, where the environment field is
averaged away. The coarse-grained (closed time-path) effective action
(Hu and Zhang, 1990; Hu, 1991b; Sinha and Hu, 1991)  is intimately related to
the influence action (Feynman and Vernon, 1963) which is needed for a
full display of backreaction effects in quantum statistical systems
(Hu, 1991a, 1993).
The coarse-grained effective action has a real part which is responsible
for particle production, while the influence action has also an imaginary part
responsible for noise. The equation of motion derived from the influence action
is the master equation for the reduced density matrix of the system after
details of the environment are traced out.
In the semiclassical limit the Wigner function associated with the reduced
density matrix obeys the Fokker-Planck equation, while, equivalently,
the system variable obeys a Langevin equation with an explicit noise
term whose distribution function depends on the nature of  and
the system's coupling with the environment.
This extended formalism in terms of the influence functional
provides a more complete platform for
the discussion of both dissipation and fluctuation processes.

     Many physical processes in the macroscopic
world manifest dissipative behavior, which is time-asymmetric. This is
contradictory to the basic laws governing the microscopic world, which
is time-symmetric. To resolve this difference is one of
the central tasks of statistical mechanics. One way is to conceive of
a natural transformation (or spontaneous evolution, see Calzetta and Hu, 1993a)
of a closed system to an open system involving the procedures
outlined in  Example C, i.e.,
separation of the system (the relevant variables) from the environment
(the irrelevant variables), choice of boundary conditions, and
averaging (coarse-graining) of the irrelevant variables.
Backreaction of the averaged effect of the irrevelant variables modifies
the dynamics of the relevant variables with a dissipative contribution.
It is through this means that random microscopic reversible processes
can bring forth irreversible behavior in the systematic macroscopic
dynamics. The connection between these two aspects is best captured in the
fluctuation-dissipation (FD) relation. In a concrete form, it provides a
microscopic derivation of the kinetic coefficients (e.g. viscosity function).
It is also one of the means that
the quantum world described by wave functions and interference
effects can be related to the classical world described by the classical
equations of motion. We will discuss the meaning of the fluctuation-dissipation
relation and the environment-induced decoherence effect here (Zurek, 1981;
Joos and Zeh, 1985; Zeh, 1986), but leave the discussion of
its relation with noise and classical structure elsewhere
(Hu, 1991a, 1993; Gell-Mann and Hartle, 1993). They are interrelated.

     The FD relation is often written for equilibrium (finite temperature $T$)
conditions and derived via linear-response theories (Callen and Welton, 1951;
Kubo, 1959). We believe that,
owing to its general nature, a relation should exist for non-equilibrium,
and for quantum ($T=0$) processes. In a recent work (Hu, Paz and Zhang, 1992,
1993a; see also Sinha and Sorkin, 1992)
we have proven at least the latter case in quantum Brownian motion
models. This provides the theoretical basis for a statistical interpretation
of quantum backreaction processes, which include the well-known
radiation-reaction problem in electrodynamics, as well as the
backreaction problems in semiclassical cosmology.
For the purpose of extending the fluctuation -dissipation relation to
quantum fields, we used path-integral methods.
Our results are summarized as follows.

Consider a Brownian particle with mass $M$ interacting with a thermal bath at
temperature $T=(k_B\beta)^{-1}$.
The classical action of the Brownian particle is
\begin{equation}
S_S[x]=\int_0^tds\Bigl\{{1\over 2}m\dot x^2-V(x)\Bigr\}
\end{equation}
The bath consists of a set of harmonic oscillators with mass $m_n$ whose
motion is described by the classical action
\begin{equation}
S_E[\{q_n\}]
=\int_0^tds\sum_n\Bigl\{
 {1\over 2}m_n\dot q_n^2
-{1\over 2}m_n\omega^2_nq_n^2 \Bigr\}
\end{equation}
Assume as an example that the system and environment interacts via a
biquadratic coupling with action
\begin{equation}
S_{int}[x,\{q_n\}]
=\int\limits_0^tds\sum_n\Bigl\{-\lambda C_nx^2q_n^2 \Bigr\}
\end{equation}
Here $\lambda$ is a coupling constant multiplied to each $ C_n$ which is
assumed to be small for perturbation calculations. The case of linear
coupling has been derived by many authors (e.g., Feynman and Vernon, 1963;
Caldeira and Leggett, 1983; Unruh and Zurek, 1989; Hu, Paz and Zhang, 1992).
For biquadratic coupling, the fluctuation-dissipation relation between
the second noise kernel $ \tilde \nu(s_1-s_2) $ and the dissipation kernel
$ \gamma(s_1-s_2) $ can be written down explicitly as
(Hu, Paz and Zhang, 1993a)

\begin{equation}
\hbar\tilde \nu(s)
=\int\limits_0^{+\infty}ds'~K(s-s')~\gamma(s')
\end{equation}

\noindent where the time convolution kernel $ K(s) $ is given by

\begin{equation}
K(s)
=\hbar\int\limits_0^{+\infty}{d\omega\over\pi}~
 \biggl\{{1+\coth^2{1\over 4}\beta\hbar\omega
 \over 2\coth{1\over 4}\beta\hbar\omega }\biggr\}~
 \omega\cos\omega s
\end{equation}

\noindent
except for the temperature dependent factor (the term within the curly
brackets)
   ,
this has the same form as the linear coupling case
(which is given by $coth(\beta \hbar \omega/2)$).
For higher order couplings with an action $\lambda C_n f(x) q^r_n$
the FD relation again has the same form as (14) and (15), only that
the temperature-dependent factor is different.
In the high temperature limit $ k_BT\gg \hbar\Gamma $, where $ \Gamma $ is the
cutoff frequency of the bath oscillators,
$ K(s)=2k_BT\delta(s)$ and the fluctuation-dissipation relation reduces to
\begin{equation}
\hbar\tilde \nu(s_1-s_2)=2k_BT\gamma(s_1-s_2)
\end{equation}

\noindent which is the famous Einstein formula (Einstein, 1905).

In the zero temperature limit $~\beta\hbar\omega\to {+\infty},~ $
\begin{equation}
K(s)
=\hbar\int\limits_0^{+\infty}
 {d\omega\over\pi}~\omega\cos\omega s
\end{equation}
It is interesting to note that the fluctuation-dissipation relations for
the linear and the nonlinear coupling models we have studied are identical
both in the high temperature and the zero temperature limits. This
insensitiveness to the different system-bath couplings reflects that it is a
categorical relation (backreaction) between the stochastic stimuli
(fluctuation-noise) of the environment and the averaged response of a system
(dissipation-relaxation) which has a much deeper
and broader meaning than that associated with the special cases studied in
the literature.
We have also derived the influence action for field theory models with
non-linear coupling and colored noise environments (Zhang, 1990; Hu, Paz and
Zhang, 1993b) and found that a set
of FD relations exist which are identical in form to the quantum mechanical
results given above. This seems to confirm our earlier suggestion
about the universality of such relations (Hu, 1989).
The FD relation suggests how macroscopic irreversibility can arise
from microscopic reversible processes. It is in this capacity that it is
relevant to the time-asymmetry problem.

The extension of the quantum Brownian motion results to quantum cosmology
is under investigation. This requires first an upgrading in the treatment of
the
cosmological backreaction problem described above
from the semiclassical to the full quantum level (describing wave functions
of the Universe). One also needs to
generalize this problem to statistical ensembles (of quantum states of
the Universe) and study the evolution of the  reduced density matrix
of the Universe obtained by tracing out, say, the scalar fields viewed as
the environment variables. (See, e.g., Paz and Sinha, 1991, 1992)
Consideration of this cosmological backreaction problem
in the statistical context pushes the domain of validity of the
fluctuation-dissipation relation to a new level, that which involves
fluctuations of quantum fields and dissipative spacetime dynamics (Hu
and Sinha, 1993a).
This relation viewed in the cosmological context has direct implications
on the notion of gravitational entropy and the time-asymmetry
issue, as we now show.

\section{ Coarse-Graining and Dissipation in Spacetime: Example in
          minisuperspace cosmology}

In a statistical-mechanical interpretation of the problem of backreaction due
to particle creation, the background spacetime plays the role of the system
while the scalar field that of the environment. The backreaction can be
calculated by the effective action method in loop expansions. In the ordinary
approach, a background-fluctuation field decomposition is assumed, and the
backreaction is due to the radiative correction effects $ O(\hbar)$ of the
matter field like vacuum fluctuation and particle creation. One can generalize
this method to treat quantum statistical processes involving coarse-graining.
Suppose one separates the field of the combined system  $\phi$ into two parts:
the system field $\bar \phi$ and the environment field $\tilde \phi$, i.e.,
$ \phi= \bar \phi + \tilde \phi$, and assumes that they are coupled weakly
with a small parameter $\lambda$. One can then construct a coarse-grained
effective action $ \Gamma[\bar\phi]$ by integrating away the environment
variables. This procedure has been used in a renormalization
group theory treatment of critical phenomena in the inflationary Universe
(Hu and Zhang, 1990, Hu, 1991b). For quantum cosmology, one can use this method
to study the effect of truncation in the gravitational degrees of freedom,
and discuss the validity of the minisuperspace approximation. (A more
comprehensive discussion of viewing minisuperspace as a quantum open system
in quantum cosmology is given in Hu, Paz and Sinha, 1993)

\subsection{Minisuperspace Approximation}

Those cosmological models most often
studied, like the Robertson-Walker, de Sitter, and the Bianchi universes,
which possess high symmetries are but a small class of a large set of possible
cosmological solutions of the Einstein equations. In terms of superspace,
the space of all three-geometries, (Wheeler, 1968; DeWitt, 1967)
these are the lower-dimensional minisuperspaces (Misner, 1972)
(e.g., the mixmaster Universe with parameters $\alpha,
\beta_+, \beta_- $ is a three-dimensional minisuperspace). In quantizing
just the few lowest modes, as is often done in quantum cosmology studies,
one ignores by fiat all these other modes. Is the minisuperspace quantization
justified? (Kuchar and Ryan, 1986, 1989) Under what
conditions is it justified?  What is the backreaction effect
of the inhomogeneous modes on the homogeneous mode?
One can view the homogeneous geometry as the system and the matter fields (or
the inhomogeneous perturbations of spacetime, the gravitons) as the
environment, and use the coarse-grained effective action to calculate
the averaged effect of the environment on the system.
Notice the similarity with the statistical mechanical problems we have
treated above. In one illustrative calculation (Sinha and Hu, 1991) we used
a model of self-interacting quantum fields to mimic the nonlinear coupling
of the gravitational waves modes (WKB time is used as it
provides correct semiclassical results) and obtained an
effective Wheeler-DeWitt equation for the minisuperspace sector with
a new term containing a nonlocal kernel. Similar in form to
Eq.(7) in the particle creation backreaction problem,
it signifies the appearance of dissipative effects in the dynamics of the
minisuperspace variables due to
their interaction with the inhomogeneous modes. Thus one can conclude
that the minisuperspace approximation is valid only if this dissipation
is small. In the same sense as the other statistical processes we have
considered above, the appearance of dissipation creates
an arrow of time in the minisuperspace sector. This also provides one way
to define gravitational entropy. Notice that in this view,
as long as one limits one's
observation to a subset of all possible geometrodynamics, and allows for some
special initial conditions,  dissipative
behavior and the emergent arrow of time are unavoidable consequences.

\subsection{ Gravitational Entropy}

The entropy of gravitational fields has been studied in connection with
self-gravitating matter (Lynden-Bell and Wood, 1967; Lynden-Bell and
Lynden-Bell, 1977; Sorkin, Wald and Zhang, 1984),
with black holes (Hawking, 1975; Sorkin, 1986),
with cosmology (Penrose, 1979) and with gravitons (Smolin 1984). We shall
consider it for cosmological spacetimes without event horizon
(See Davies, Ford and Page, 1989 for the case of de Sitter universe, which has
an event horizon). Gravitational
entropy of the Universe has
also been discussed before in conjunction with quantum dissipative processes
in the early Universe (Hu, 1983, 1984). Here I want to discuss it in the
context of quantum cosmology (see also Kandrup, 1989)
and the theme of the present conference, time-asymmetry.

Following the idea of minisuperspace approximation in quantum cosmology
discussed above as a backreaction problem and generalizing the wave functions
of the Universe to density matrices of the Universe, we can work with the
reduced density matrix of the Universe  constructed by tracing out the
matter fields or the higher gravitational modes  and define
a gravitational entropy of the homogeneous Universe as
\begin{equation}
S= -Tr \rho_{red} ln \rho_{red}                         
\end{equation}
{}From the theory of subdynamics, we know that $S$ increases with
time. (Note again that some notion of time has to be introduced beforehand,
e.g., the WKB time, or the 4-volume time of Sorkin, 1993).
The arrow of time arises as the direction of
information flow from the relevant (spacetime, or the homogeneous
gravitational modes) to the irrelevant (the matter fields, or
the inhomogeneous modes) degrees of freedom. (See Hu, 1993; Hu and Sinha, 1993b
for details.)\\

     In this and earlier sections I have only sketched the statistical
nature of certain quantum processes in semiclassical gravity and
quantum cosmology, but I hope this array of examples and questions
-- from billiard balls to ink drops to plasma waves to particle creation
to anisotropy damping to density matrix of the Universe --
has demonstrated to you, despite the great disparity of their context,
the universality of the issues involved and the conceptual unity in our
understanding.

Up to now I have only discussed HOW one can see dissipation and the arrow
of time arising in the system from coarse-graining the environment.
I have not mentioned anything about the more fundamental and difficult
questions in the system-environment approach to these issues
in statistical mechanics, i.e.,
WHY? Why should the system be regarded as such?
Why should the separation be made as such?
Why should a sector be viewed as the system and get preferential
treatment over the others. The answer to these questions when
raised in the cosmological context can be more meaningfully sought with
the open-system perspective if the spacetime has some distinguished
global or physical structures like event horizon, particle horizon, non-trivial
topology, etc. One can then define an objectively meaningful domain
for the system and study its effective dynamics. The outcome also
depends on how the coarse-graining (measurement, observation, participation)
is taken, and how effective it is in producing persistent robust structures
(Woo, 1989). Consistency in the behavior of the system after these procedures
are taken (such as how stable any level of structure is with respect to
iteration of the same coarse-graining routines, and
how sensitive the open system is with respect to variations of coarse-graining)
is certainly an important criterion in any consideration.
I will now say a few things on these issues to conclude my talk.

\section{ Coarse-Graining and Persistent Structure in the Physical World}

      Let me summarize the main points of this talk and suggest
a few questions to explore on the issue of irreversibility in cosmology.

     On the whole, there are two different causes for the appearance of
irreversibility: one due to special initial conditions,
the other due to dissipation. \footnote {As discussed earlier, dissipation
also requires the stipulation of a somewhat special initial condition, i.e.,
that the system is not in an equilibrium state;
but `not more special than it needs to be' --in the words of R. Sorkin.}
The first class is {\it a priori}
determined by the initial conditions, the other is {\it a posteriori}
rather insensitive to the initial conditions.
Of the examples we have given, the first class includes chaotic dynamics,
Landau damping, vacuum particle creation, the second class includes molecular
dynamics, diffusion, particle creation with interaction, anisotropy
dissipation,
decoherence. Appearance of dissipation is accompanied by a degradation of
information via coarse graining (such as the molecular chaos assumption
in kinetic theory,
restriction to one-particle distribution in particle creation with interaction,
`integrating out' some class of histories in decoherence).
An arrow of time appears either because of some special prearranged
conditions or as a consequence of coarse-graining introduced to the system.
The issues we have touched on involve the transformation of a closed to
an open system, the relation between the microscopic and the macroscopic world,
and the transition from quantum to classical realities.
Many perceived phenomena in the observable physical world,
including the phenomenon of time-asymmetry, can indeed be understood in
the open-system viewpoint via
the approximations introduced to the objective microscopic
world by a macroscopic observer. We have discussed
the procedures which can bring about these results.
However, what to me seems more important and challenging
is to explore under what conditions the outcomes
become less subjective and less sensitive to these procedures, such as the
system- environment split and the coarse-graining of the environment.
These procedures provide one with
a viable prescription to get certain general qualitative results,
but are still not specific and robust enough to explain
how and why the variety of observed phenomena in the physical world
arise and stay in their particular ways. To address these issues one should ask
a different set of questions:\\

{\it 1) By what criteria are the system variables chosen?
   --collectivity and hierarachy of structure and interactions}

     In a model problem, one picks out the system variables --
be it the Brownian particle or the minisuperspace variables -- by fiat.
One defines one's system in a
particular way because one wants to calculate the properties of that particular
system.  But in the real world, certain
variables distinguish themselves from others because they possess a
relatively well-defined, stable, and meaningful set of properties for which
the observer can carry out measurements and derive meaningful results. Its
meaningfulness is defined by the range of validity or degree of precision or
the level of relevance to what the observer chooses to extract information
from.
In this sense, it clearly carries a certain degree of subjectivity--
not in the sense of arbitrariness in the will of the observer,
but in the specification of the parameters of observation and measurement.
For example, the thermodynamic and hydrodynamic variables are only good for
systems  close to equilibrium; in other regimes one needs
to describe the system in terms of kinetic-theoretical or
statistical-mechanical
variables.
The soundness in the choice of a system in this example thus depends on the
time scale of measurement compared to the relaxation time.
As another example, contrast the variables used in nuclear collective model
and the independent nucleon models. One can use the rotational-vibrational
degrees of freedom to depict some macroscopic properties of the motion of
the nucleus, and one can carry out meaningful calculations of the
dissipation of the collective trajectories (in the phase space of the nucleons)
due to stochastic forces. In such cases, the non-collective degrees of freedom
can be taken as the noise source. However, if one is interested in how the
independent nucleons contribute to the properties of the nucleus,
such as the shell structure, one's system variable should, barring some simple
cases, not be the elements of the $SO(3)$ group, or the $SU(6)$ group.
At a higher still energy where the attributes of the quarks and the gluons
become apprarent, the system variables for the calculation of, say, the
stability of the quark-gluon plasma should change accordingly.  The level of
relevance which defines one's system changes with the level of structure
of matter and the relative importance of the forces at work at that level.
The improvement of the Weinberg-Salam model with $W, Z$ intermediate bosons
over the Fermi model of four point interactions is what is needed in probing
a deeper level of interaction and structure which puts
the electromagnetic and weak forces on the same footing.
Therefore, one needs to
explore the rules for the formation of such relatively distinct and stable
levels, before one can sensibly define one's system (and the environment)
to carry out meaningful inquiries of a statistical nature.

What is interesting here
is that these levels of structures and interactions come in approximate
hierarchical order (so one doesn't need QCD to calculate the rotational
spectrum
of a nucleus, and the Einstein spacetime manifold picture will hopefully
provide most of what we need in the post-Planckian era). One needs both
some knowledge of the hierarchy of interactions (e.g., Weinberg 1974)
and the way effective theories emerge-- from `integrating out' variables at
very different energy scales in the hierarchical structure
(e.g., ordinary gravity plus grand-unified
theory regarded as a low energy effective Kaluza-Klein theory)
The first part involves fundamental constituents and interactions and
the second part the application of statistical methods.
One should also keep in mind that what is viewed as fundamental at one level
can be a composite or statistical mixture at a finer level.
There are system-environment separation schemes which are designed to
accomodate or reflect these more intricate structures, such as the
mean field-fluctuation field split, the dynamics of correlations (Balescu,
1975;
Calzetta and Hu, 1988) and the multiple source formalism
(Cornwall, Jackiw and Tomboulis 1974; Calzetta and Hu, 1993a). The
validity of these approximations depends on where exactly one wants to probe
in between any two levels of structure. Statistical properties of the
system such as the appearance of
dissipative effects and the associated irreversibility character of the
dynamics in an open system certainly depend on this separation.\\

{\it 2)  How does the behavior of the subsystem depend on coarse-graining?--
    sensitivity and variability of coarse-graining,
    stability and robustness of structure}

    Does there exist a common asymptotic regime as the result of including
successively higher order iterations in the same coarse-graining routine?
This measures the sensitivity of the end result to a particular kind of
coarse-graining. How well can different kinds of coarse-graining measure
produce and preserve the same result? This is measured by its variability.
Based on these properties of coarse-graining, one can discuss the relative
stability of the behavior of the resultant open system
after a sequence of coarse-grainings within the same routine,
and its robustness with respect to changes
to slightly different coarse-graining routines.

    Let me use some simple examples to illustrate what this problem is about.
When we present a microscopic derivation of the transport coefficients
(viscosity, heat conductivity, etc) in kinetic theory via the
system-environment
scheme, we usually get the same correct answer independent of the way
the environment is chosen or coarse-grained. Have we ever wondered why?
It turns out that this is the case only if we
operate in the linear-response regime. (Feynman and Vernon 1963). The
linear coupling between the system and the environment makes this dependence
simple. This is something we usually take for granted, but has some deeper
meaning. For nonlinear coupling, the above problem becomes nontrivial.
Another aspect of this problem can be brought out in the following
consideration (Balian and Veneroni, 1987).
Compare these two levels of structure and interaction:
hydrodynamic regime and kinetic regime. Construct the relevant entropy
(in the information theory sense) from the one-particle distribution $\rho$
under the constraint that the average of any physical variable $O$ is given
by $ <O>= Tr \rho O $. $\rho$ changes with different levels of
coarse-graining. In terms of the one-particle classical distribution function
$f_1$ the entropy function $S$ is given by

\begin{equation}
S_B = \int d\vec{r}d\vec{p} f_1(\vec{r},\vec{p})[1 - lnh^3
f_1(\vec{r},\vec{p})]
\end{equation}
in Botzmann's kinetic theories, and
\begin{equation}
S_H \sim N^3,~~~  N = \int d\vec{r} d\vec{p} f_1(\vec{r}, \vec{p})
\end{equation}
in hydrodynamics. Notice that $S_H > S_B$  is a maximum in the sequence of
different coarse-graining procedures. In the terminology we introduced above,
by comparison with the other regimes, the
hydrodynamic regime is  more robust in its structure and interactions
with respect to varying levels of coarse-graining. The reason for this is,
as we know, because the hydrodynamic variables describle systems in
equilibrium.
Further coarse-graining on these systems is expected to produce the same
result. Therefore, a kind of `maximal entropy principle'
with respect to variability of coarse-graining is one way where
thermodynamically robust systems can be located.

While including successively  higher orders of the
same coarse-graining measure usually gives rise to quantitative differences
(if there is a convergent result, that is, but this condition is not
gauranteed, especially if phase transition intervenes),
coarse-graining of a different nature will in general result in very
different behavior in the dynamics of the open system.
Let us look further at the relation of variability of coarse-graining
and robustness of structure.

    Sometimes the stability of a system with respect to coarse-graining
is an implicit criterion behind the proper choice of a system. For example,
Boltzmann's equation governing the one-particle distribution function which
gives a very adequate depiction of the physical world is, as we have seen,
only the lowest order equation in an infinite (BBGKY) hierarchy.
If coarse-graining is by the order of the hierarchy -- e.g.,
if the second and higher order correlations are ignored, then one can
calculate without ambiguity the error introduced by such a truncation.
The dynamics of the open system
which includes dissipation effects and irreversible behavior will change
little as one coarse-grains further to higher and higher order
(if the series convergences, see, e.g., Dorfman, 1981).
In another approximation, for a binary gas of large mass
discrepancy, if one considers the system as the heavy mass particles,
ignore their mutual interactions and coarse-grain the effect of the light
molecules on the heavy ones, one can turn the Boltzmann equation into a
Fokker-Planck equation for Brownian motion, and get qualitatively
very different results in the behavior of the system.

In general the variability of different coarse-grainings in producing a
qualitatively similar result is higher (more variations allowed) when
the system one works with is closer to a stable level in the interaction
range or in the hierarchical order of structure of matter.
The result is more sensitive to different coarse-graining measures
if it is far away from a stable structure,
usually falling in between two stable levels.

One tentative analogy may help to fix these concepts:
robust systems are like
the stable fixed points in a parameter space in the renormalization group
theory description of critical phenomena: the points in a trajectory are
the results of performing successive orders of the same coarse-graining
routine on the system (e.g., the Kadanoff-Migdal scaling),
a trajectory will form if the coarse graining  routine is stable. An
unstable routine will produce in the most radical situations
a random set of points.
Different trajectories arise from different
coarse-graining routines. Neighboring trajectories will converge if the
system is robust, and diverge if not. Therefore the existence of a stable
fixed point where trajectories converge to is an indication that the
system is robust. Only robust systems survive in nature and carry definite
meaning in terms of their persistent structure and systematic evolutions.
This is where the relation of coarse-graining and persistent structures
enter.

So far we have only discussed the activity around one level of robust
structure.
To investigate the domain lying in-between two levels of structures
(e.g., between nucleons and quark-gluons) one needs to first know the basic
constituents and interactions of the two levels.  This brings back our
consideration of levels of structures above. Studies in the properties
of coarse-graining can provide a useful guide to venture into the
often nebulous and evasive area between the two levels
and extract meaningful results pertaining
to the collective behavior of the underlying structure.
But one probably cannot gain new information about the fine
structure and the new interactions from the old just by these statistical
measures. (cf. the old bootstrapping idea in particle physics versus the
quark model). In this sense, one should not expect to gain
new fundamental information about quantum gravity just by extrapolating
what we know about the semiclasical theory, although studying the way how the
semiclassical theory takes shape (viewed as an effective theory ) from a
more basic quantum theory is useful.
It may also be sufficient for what we can
understand or care about in this later stage of the Universe we now live in.

     There are immediate consequences from these theoretical discussions for
cosmology. Questions like, why the Universe
should in its later stage settle into the highly symmetric state of
isotropy and homogeneity? Is this a particular choice of the `system' from
the beginning, or is it a consequence of coarse-graining an initial larger set
of possibilities both in the spacetime and the matter degrees of freedom?
What are the stable coarse-graining routines? How different can the
coarse-graining routines be to still produce robust results?
I have just begun to explore these questions in a number of ways. They are,
\noindent a) Viewing the homogeneus cosmology as the infrared sector of
spacetime excitations, and using the rules of dimensional
reduction as possible explanation for its prevalance. (Hu, 1990)
\noindent b) Gravity as an effective theory and geometric structure
as collective degrees of freedom (Sahkarov, 1968; Adler and Zee, 1984).
\noindent c) Einstein's gravity as the hydrodynamic limit of a
nonlinear and nonlocal theory, drawing on the insight from the
behavior of the Boltzmann equation, the BBGKY hiearachy and the
long-wavelength hydrodynamic approximations.
There is no time to describe them here, but I hope the discussions
on the properties and origins of irreversible processes
in cosmology, sketchy as they have been presented here, can help us gain a
better perspective of
the universality of these issues in physics and provide some theoretical basis
for further discussions of their meanings.

\section{Acknowledgements}

I thank Raphael Sorkin for a careful and thoughtful reading of this manuscript
and for evoking many interesting discussions on various issues raised in this
paper. Research is supported in part by the National Science Foundation under
grant PHY91-19726.

\section{Questions and Comments}
\it Cover: \rm In his question, P. Davies has suggested that the
entropy of a gravitational   field might be replaced by Kolmogorov
(or algorithmic)
complexity.  It should be noted that entropy, as well as algorithmic
complexity, are descriptive complexities.  Moreover, they usually
agree.  And in the special case of equipartition of energy (or
probability), entropy and Kolmogorov complexity equal the logarithm
bof the number of microstates of the given macrostate.

\noindent
\it Hartle: \rm If we are going to consider complexity then we are going
to have to ask ``whose complexity is it'' that is, what coarse-graining
is going to be used to compute it?

\noindent \it Hu: \rm Although coarse-graining has a strong element
of subjectivity, those classes which lead to physical reality
(including complexity) which is agreed upon by a large class of
observers (including us) merit special attention.  It is important to study
the {\it criteria} and {\it conditions} for these coarse-grainings to
be favorably selected in the evolutionary process which give rise to
persistent structures (persistent at least to the degree we can perceive them).

\section{References}

\begin{itemize}
\begin{description}
\item[] Adler, S. L. (1982) Rev. Mod. Phys., {\bf 54}, 719.
\item[] Albrecht, A.  and Steinhardt, P. J. (1982) Phys. Rev. Lett., {\bf
48}, 1220.
\item[] Ashtekar, A.  and Stachel, J. (1991) eds {\it Conceptual Problems in
Quantum Gravity} (Birkhauser, Boston)
\item[] Balescu, R. (1975) \it Equilibrium and Nonequilibrium Statistical
Mechanics, \rm (Wiley, New York).
\item[] Balian, R. and Veneroni, M. (1987) Ann. Phys. (N.Y.), {\bf 174},
229-244.
\item[] Barrow, J. D. (1982) Phys. Rep., {\bf 85}, 1.
\item[] Bekenstein, J. D. (1973) Phys. Rev., {\bf D7}, 2333.
\item[] Bekenstein, J.D. (1974) Phys. Rev., {\bf D9}, 3292.
\item[] Berger, B. K. (1992) in Proc. GR13, Cordoba, Argentina
\item[] Bogoiavlenskii, O. I. (1985) \it Methods in the Qualitative Theory of
Dynamical Systems in Astrophysics and Gas Dynamics \rm (Springer-Verlag,
Berlin).
\item[] Caldeira, A. O. and Leggett, A. J. (1983) Physica, {\bf A121}, 587.
\item[] Callen, H. B. and Welton, T. A. (1951)  Phys. Rev. {\bf 83}, 34.
\item[] Calzetta, E. (1989) Class. Quantum Grav., {\bf 6}, L227.
\item[] Calzetta, E. (1991) Phys. Rev., {\bf D43}, 2498.
\item[] Calzetta, E. and Hu, B. L. (1987) Phys. Rev., {\bf D35}, 495.
\item[] Calzetta, E. and Hu, B. L. (1988) Phys. Rev., {\bf D37}, 2878.
\item[] Calzetta, E. and Hu, B. L. (1989) Phys. Rev., {\bf D40}, 656.
\item[] Calzetta, E. and Hu, B. L. (1993a) ``Decoherence of Correlation
Historie
   s"
in {\it Directions in General Relativity} Vol 2 (Brill Festschrift)
eds. B. L. Hu and T. A. Jacobson (Cambridge Univ., Cambridge)
\item[] Calzetta, E. and Hu, B. L. (1993b) ``From Kinetic Theory to Brownian
Mot
   ion"
unpublished.
\item[] Calzetta, E. and Mazzitelli, F. (1991) Phys. Rev., {\bf D42}, 4066.
\item[] Candelas, P. and Sciama, D. W. (1977) Phys. Rev. Lett., {\bf38}, 1372.
\item[] Coleman, S., Hartle, J.,  Piran, T., and Weinberg, S. (1990) eds
\it Quantum Cosmology  and Baby Universes \rm (World Scientific, Singapore).
\item[] Cornwall, J. M., Jackiw, R., and Tomboulis, E. (1974) Phys. Rev.
 {\bf D10}, 2428.
\item[] Davies, P. C. W. (1975) J. Phys. {\bf A8}, 609.
\item[] Davies, P. C. W., Ford L. and Page, D. (1987) Phys. Rev., {\bf
D34}, 1700.
\item[] De Groot, S. R., van Leeuwen, W. A. and van Weert, Ch. G. (1980)
\it Relativistic Kinetic Theory \rm (North-Holland, Amsterdam)
\item[] DeWitt, B. S. (1967) Phys. Rev., {\bf 160}, 1113.
\item[] Dorfman, R. (1981) in \it Perspectives in
Statistical Physics \rm Vol. IX,  Eds. H.J. Raveche (North-Holland, Amsterdam).
\item[] Einstein, A. (1905) Ann. Phys. (Leipzig) {\bf 17}, 549
\item[] Feynman, R. P. and Vernon, F. L. (1963) Ann. Phys. (N.Y.),
{\bf 24}, 118.
\item[] Ford, G. W., Kac, M. and Mazur, P. (1963) J. Math. Phys. {\bf 6}, 504
\item[] Gell-Mann, M. and Hartle, J. B. (1990) in \it Complexity, Entropy
and the Physics of Information \rm ed. W. H. Zurek (Addison-Wesley, N.Y.).
\item[] Gell-Mann, M. and Hartle, J. B. (1993) Phys. Rev., {\bf 47}
\item[] Goldwirth, D. S. (1991) Phys. Lett. {\bf B256}, 354.
\item[] Grabert, H. (1982) \it Projection Operator Techniques in
Nonequilibrium Statistical Mechanics \rm (Springer Verlag, Berlin).
\item[] Grabert H., Schramm, P. and Ingold, G. (1988) Phys. Rep., {\bf 168},
115
   .
\item[] Guth, A., (1981) Phys. Rev., {\bf D23}, 347.
\item[] Halliwell, J. J. (1993) \it Quantum Cosmology \rm (Cambridge
Univ. Press, Cambridge).
\item[] Hartle, J. B. (1983) in \it The Very Early Universe, \rm eds.
G. Gibbons, S. W. Hawking and S. Siklos (Cambridge Univ. Press, Cambridge)
\item[] Hartle, J. B. and Hawking, S. W. (1983) Phys. Rev., {\bf D28}, 2960.
\item[] Hartle, J. B. and Hu, B. L. (1980) Phys. Rev., {\bf D21}, 2756.
\item[] Hawking, S. W. (1975) Commun. Math. Phys., {\bf 87}, 395.
\item[] Heller, J. P. (1960) Am. J. Phys., {\bf 28}, 348-353.
\item[] Hu, B. L. (1982) in \it  Proc. Second Marcel Grossmann Meeting 1979,
\rm Ed. R. Ruffini, (North-Holland, Amsterdam).
\item[] Hu, B. L. (1983) Phys. Lett., {\bf 97A}, 368.
\item[] Hu, B. L. (1984) in \it Cosmology of the Early Universe,
\rm Ed. L. Z. Fang and R. Ruffini (World Scientific, Singapore).
\item[] Hu, B. L. (1987) ``Recent Development in Cosmological Theories",
IAS Preprint, Princeton, IASSNS-HEP87/15.
\item[] Hu, B. L. (1989) Physica, {\bf A158}, 399.
\item[] Hu, B. L. (1990) ``Quantum and Statistical Effects in Superspace
Cosmology" in {\it Quantum Mechanics in Curved Spacetime}, ed. J. Audretsch
and V. de Sabbata (Plenum, London, 1990)
\item[] Hu, B. L. (1991a) ``Statistical Mechanics and Quantum Cosmology",
in {\it Proc. Second International Workshop on Thermal Fields and Their
Applications}, eds. H. Ezawa et al (North-Holland, Amsterdam, 1991)
\item[] Hu, B. L. (1991b) ``Coarse-Graining and Backreaction in Inflationary
and Minisuperspace Cosmology" in {\it Relativity and Gravitation: Classical
and Quantum}, Proc. SILARG VII, Cocoyoc, Mexico 1990,
eds. J. C. D' Olivo et al (World Scientific, Singapore, 1991)
\item[] Hu, B. L. (1993) ``Quantum Statistical Processes in the Early Universe"
in {\it Quantum Physics and the Universe}, Proc. Waseda Conference, Aug. 1992
ed. M. Namiki, K. Maeda, et al  (Pergamon Press, Tokyo, 1993)
\item[] Hu, B. L. and Kandrup, H. E. (1987) Phys. Rev., {\bf D35}, 1776.
\item[] Hu, B. L. and Parker, L. (1978) Phys. Rev., {\bf D17}, 933.
\item[] Hu, B. L. and Pavon, D. (1986) Phys. Lett., {\bf180B}, 329.
\item[] Hu, B. L., Paz, J. P., and Sinha, S. (1993)
``Minisuperspace as a Quantum Open System" in
{\it Directions in General Relativity}  Vol. 1, (Misner Festschrift)
eds B. L. Hu, M. P. Ryan and C. V. Vishveswara (Cambridge Univ., Cambridge)
\item[] Hu, B. L., Paz, J. P. and  Zhang, Y. (1992) Phys. Rev.,
{\bf D45}, 2843.
\item[] Hu, B. L., Paz, J. P. and  Zhang, Y. (1993a)
``Quantum Brownian Motion in a General Environment
II. Nonlinear coupling and perturbative approach'' Phys. Rev. {\bf D47}  (1993)
\item[] Hu, B. L., Paz, J. P. and  Zhang, Y. (1993b)
``Stochastic Dynamics of Interacting Quantum Fields'' Phys. Rev. D (1993)
\item[] Hu, B. L., Paz, J. P. and  Zhang, Y. (1993c)
``Quantum Origin of Noise and Fluctuation in Cosmology"
in {\it Proc. Conference on the Origin of Structure in the Universe}
Chateau du Pont d'Oye, Belgium, April, 1992, ed. E. Gunzig and P. Nardone
(NATO ASI Series) (Kluwer, Dordrecht, 1993)
\item[] Hu, B. L. and Sinha, Sukanya (1993a)
"Fluctuation-Dissipation Relation in Cosmology"    Univ. Maryland preprint
\item[] Hu, B. L. and Sinha, Sukanya (1993b)
"Spacetime Coarse-Graining and Gravitaional Entropy" Univ. Maryland preprint
\item[] Hu, B. L. and Zhang, Y. (1990)  ``Coarse-Graining, Scaling, and
Inflation" Univ. Maryland Preprint 90-186 (1990)
\item[] Hu, B. L. and Zhang, Y. (1992)  ``Uncertainty Principle
at Finite Temperature"  Univ. Maryland preprint (1992)
\item[] Hu, B. L. and Zhang, Y. (1993) ``Quantum and Thermal Fluctuations,
Uncertainty Principle, Decoherence and Classicality" in {\it Quantum Dynamics
of Chaotic Systems}: Proc. Third International
Workshop on Quantum Nonintegrability, Drexel University, Philadelphia,
May 1992, ed. J. M. Yuan, D. H. Feng, and G. M. Zaslavsky
(Gordon and Breach, Langhorne, 1993)
\item[] Isham, C. J. (1991) "Conceptual and Geometrical Problems in Quantum
Gravity" Lectures at the Schladming Winter School, Imperial College preprint
TP/90-91/14
\item[] Joos, E. and Zeh, H. D. (1985) Z. Phys. {\bf B59}, 223.
\item[] Kandrup, H. E. (1988) Phys. Rev. {\bf D37}, 3505.
\item[] Kandrup, H. E. (1988) Class. Quanatum Grav. {\bf 5}, 903
\item[] Keldysh, L. V. (1964) Zh. Eksp. Teor. Fiz. {\bf 47}, 1515
[Sov. Phys. JEPT {\bf 20}, 1018 (1965)]
\item[] Kreuzer, H. J. (1981) \it Nonequilibrium Thermodynamics and Its
Statistical Foundations \rm(Oxford Univ., Oxford).
\item[] Kubo, R. (1959) {\it Lectures in Theoretical Physics}, Vol 1,
pp 120-203 (Interscience, N. Y. 1959)
\item[] Kuchar, K. (1992) ``Time and Interpretaions in Quantum Gravity"  in
{\it Proceedings of the 4th Canadian Conference on General Relativity and
Relativistic Astrophysics}, eds. G. Kunstatter, D. Vincent and J. Williams
(World Scientific, Singapore).
\item[] Kuchar, K. and Ryan, M. P., Jr., (1986)  In \it Gravitational Collapse
and Relativity \rm Ed. H. Sato and T. Nakamura
(World Scientific, Singapore).
\item[] Kuchar, K. and Ryan, M. P. Jr., (1989) Phys. Rev., {\bf D40},
3982.
\item[] Linde, A. (1982) Phys. Lett., {\bf 108B}, 389.
\item[] Lynden-Bell D. and Wood, R. (1967) MNRAS {\bf 136}, 101.
\item[] Lynden-Bell D. and Lynden-Bell, R. M. (1977) MNRAS {\bf181}, 405.
\item[] Ma, S. K. (1985) \it Statistical Mechanics \rm (World Scientific,
Singapore).
\item[] Misner, C. W. (1969) Phys. Rev. Lett., {\bf 22}, 1071.
\item[] Misner, C. W. (1972) in \it Magic Without Magic, \rm Ed. J. Klauder
(Freeman, San Francisco).
\item[] Morikawa, M. (1989) Phys. Rev., {\bf D40}, 4023.
\item[] Mottola, E. (1986) Phys. Rev., {\bf D33}, 2126.
\item[] Page, D. M. (1984) private communication
\item[] Parker, L. (1969) Phys. Rev., {\bf 183}, 1057.
\item[] Parker, L. (1986) in: \it The Quantum Theory of Gravity, \rm S.
Christensen, Ed. (Adam Hilger, S. Bristol, 1986).
\item[] Paz, J.P. (1990) Phys. Rev., {\bf D40}, 1054.
\item[] Paz, J. P. and Sinha, Sukanya (1991) Phys. Rev. {\bf D44}, 1038.
\item[] Paz, J. P. and Sinha, Sukanya (1992) Phys. Rev. {\bf D45}, 2823.
\item[] Peebles, P. J.E.  (1971) \it Physical Cosmology \rm (Princeton
Univ. Press, Princeton).
\item[] Penrose, R. (1979) ``Singularities and Time-Asymmetry'' in {\em
General Relativity: an Einstein Centenary Survey}, eds. S.W. Hawking and
W. Israel (Cambridge University Press, Cambridge, 1979)
\item[] Prigogine, I. (1962) {\it Introduction to Thermodynamics of
Irreversible Processes} 2nd ed. (Wiley, New York).
\item[] Rubin, R. (1960) J. Math. Phys. {\bf 1}, 309.
\item[] Ryan, M. P., Jr., and Shepley, L.C.  (1975) \it Homogeneous
Relativistic Cosmologies \rm (Princeton Univ. Press, Princeton).
\item[] Sakharov, A. D. (1967) Dok. Akad. Nauk. SSR, {\bf 177}, 70
[Sov. Phys. Dokl {\bf12} (1968) 1040].
\item[] Sato, K. (1981) Phys. Lett. {\bf 99B}, 66.
\item[] Schwinger, J. S. (1961) J. Math. Phys. {\bf 2}, 407.
\item[] Sciama, D. W. (1979) in {\it Centenario di Einstein} Editrice
Giunti Barbara-Universitaria.
\item[] Sexl, R. U. and Urbantke, H. K. (1969) { Phys. Rev.}, {\bf 179}, 1247
\item[] Sinha, Sukanya (1991) Ph. D. Thesis, University of Maryland.
\item[] Sinha, Sukanya and Hu, B. L. (1991) Phys. Rev., {\bf D44}, 1028-1037.
\item[] Sinha, Supurna and Sorkin, R. D. (1992) Phys. Rev., {\bf B45},
8123-8126
   .
\item[] Smolin, L. (1985) Gen. Rel. Grav.,{\bf 7}, 417-437.
\item[] Sorkin, R. D. (1986) Phys. Rev. Lett. {\bf 56}, 1885-1888.
\item[] Sorkin, R. D. (1993) Int. J. Theor. Phys.
\item[] Sorkin, R. D., Wald, R. M. and Zhang, Z. J. (1981)
 Gen. Rel. Grav., {\bf 12}, 1127.
\item[] Unruh, W. G. (1976) Phys. Rev., {\bf D14}, 870.
\item[] Unruh, W. and  Zurek, W. H.  (1989) Phys. Rev. {\bf D40}, 1071.
\item[] Vilenkin, A. (1986) Phys. Rev., {\bf D33}, 3560.
\item[] Weinberg, S. (1972) \it Gravitation and Cosmology \rm (John
wiley, N.Y.).
\item[] Weinberg, S. (1980) Phys. Lett. {\bf B91}, 51
\item[] Wheeler, J. A.,(1968) in {\it Battelle Recontres}, eds. C. DeWitt
and J. A. Wheeler (Benjamin, New York).
\item[] Woo, C. H. (1989) Phys. Rev., {\bf D39}, 3174.
\item[] Zee, A. (1979) Phys. Rev. Lett., {\bf 42} 417.
\item[] Zeh, H. D. (1986) Phys. Lett. {\bf A116}, 9.
\item[] Zel'dovich, Ya. B. and Starobinsky, A. A.  (1971) Zh. Eksp.
Teor. Fiz {\bf 61}, 2161 [Sov. Phys. JETP, {\bf 34}, 1159 (1972)].
\item[] Zel'dovich, Ya. B. (1970) Pis'ma Zh Eksp. Teor. Fiz. {\bf 12}, 443
[JETP Lett., {\bf 12} (1970) 307] .
\item[] Zhang, Yuhong (1990) Ph. D. Thesis, University of Maryland.
\item[] Zurek, W. H. (1981) Phys. Rev. {\bf D24}, 1516.
\item[] Zurek, W. H. (1982) Phys. Rev. {\bf D26}, 1861.
\item[] Zwanzig, R. (1961) in \it Lectures in Theoretical Physics,
Vol. III, \rm Eds. W.E. Britten, B.W. Downes and J. Downes
(Interscience, New York).
\end{description}
\end{itemize}
\end{document}